\documentclass[conference]{IEEEtran}

\usepackage{amsmath,amsfonts}
\usepackage{array}
\usepackage[caption=false,font=normalsize,labelfont=sf,textfont=sf]{subfig}
\usepackage{textcomp}
\usepackage{stfloats}
\usepackage{url}
\usepackage{verbatim}
\usepackage{graphicx}
\usepackage{cite}
\usepackage{multirow}%
\usepackage{amsthm}%
\usepackage{mathrsfs}%
\usepackage{xcolor}%
\usepackage{textcomp}%
\usepackage{manyfoot}%
\usepackage{booktabs}%
\usepackage{listings}%
\usepackage{subcaption}

\usepackage{lmodern}
\usepackage{anyfontsize}
\usepackage{float}
\usepackage{bm}
\usepackage{mathtools}

\usepackage{tabularx,ragged2e}
\usepackage{makecell}
\newcolumntype{C}{>{\Centering\arraybackslash}X} 
\newcolumntype{P}{>{\arraybackslash}X} 
\usepackage{colortbl}

\definecolor{mygray}{gray}{0.7}
\usepackage[ruled,vlined,linesnumbered]{algorithm2e}
\DontPrintSemicolon
\SetKwComment{Comment}{$\triangleright$\ }{}
\SetKwComment{Comment}{//}{}
\SetKwFor{RepTimes}{repeat}{times}{end}
\SetKwBlock{Repeat}{repeat}{}

\usepackage{tikz}
\usetikzlibrary{positioning,shapes.geometric}

\begin{document}
\title{
Adapting Multi-Stage Distributed Real-Time Systems Design to Vehicular Edge Computing
}

\author{
\IEEEauthorblockN{Niraj Kumar, Anant Gautam}
\IEEEauthorblockA{Indian Institute Of Technology Goa, India \\
Email: \{niraj, anant2413101\}@iitgoa.ac.in}
}

\maketitle

\begin{abstract}
Vehicular Edge Computing (VEC) aims to support computation-intensive applications in intelligent transportation systems while meeting stringent latency requirements. When processed remotely, applications require access to different resource types in a specific sequence, forming a multi-stage processing pipeline. This structure closely resembles multi-stage distributed applications executing across heterogeneous resources, making established multi-stage distributed real-time scheduling techniques well suited for modeling VEC systems. The resulting model accommodates diverse resource configurations and access patterns. Building on this model, we propose a joint mapping and scheduling approach that accounts for vehicle mobility and real-time constraints. The proposed heuristic extends optimal priority assignment for multi-stage systems to the VEC setting, to construct feasible schedules while minimizing local-device energy consumption and ensuring that offloaded jobs meet their deadlines. Extensive simulations show that the proposed approach consistently outperforms both a baseline based on the same delay-composition framework and a deadline decomposition-based approach.

\end{abstract}


\IEEEpeerreviewmaketitle

\newcommand{\nk}[1]{{\color{blue}{\textbf{NK:}#1}}}
\newcommand{\tbd}[1]{{\color{purple}{\textbf{TBD}#1}}}

\newcommand{\jobInd}{\ensuremath{k}} 
\newcommand{\vehicleInd}{\ensuremath{k}} 
\newcommand{\stgInd}{\ensuremath{x}} 
\newcommand{\resInd}{\ensuremath{y}} 
\newcommand{\rsuInd}{\ensuremath{i}} 
\newcommand{\serverInd}{\ensuremath{j}} 
\newcommand{\tempInd}{\ensuremath{q}} 
\newcommand{\segInd}{\ensuremath{q}} 

\newcommand{\noRes}{\ensuremath{N^{res}}} 
\newcommand{\noSeg}{\ensuremath{N^{seg}}} 
\newcommand{\noStg}{\ensuremath{N^{stg}}} 
\newcommand{\noTasks}{\ensuremath{n}} 
\newcommand{\noJobs}{\ensuremath{N^{job}}} 

\newcommand{\job}{\ensuremath{J}}   
\newcommand{\vehicle}{\ensuremath{V}}   
\newcommand{\rsu}{\ensuremath{R}}   
\newcommand{\server}{\ensuremath{S}}   
\newcommand{\arrivalTime}{\ensuremath{a}}   
\newcommand{\deadline}{\ensuremath{d}}   
\newcommand{\dataSizeInput}{\ensuremath{\theta^{inp}}}   
\newcommand{\dataSizeResult}{\ensuremath{\theta^{res}}}   
\newcommand{\cyclesRequired}{\ensuremath{c^{req}}}   
\newcommand{\energyLocal}{\ensuremath{E^{local}}}   
\newcommand{\energyTotal}{\ensuremath{\Phi}}   
\newcommand{\position}{\ensuremath{p}}   
\newcommand{\speedVehicle}{\ensuremath{s}}   
\newcommand{\speedServer}{\ensuremath{c}}   
\newcommand{\rangeEnterTime}{\ensuremath{T^{entry}}}   
\newcommand{\rangeExitTime}{\ensuremath{T^{exit}}}   
\newcommand{\sojournTime}{\ensuremath{T^{sojourn}}}   

\newcommand{\delay}{\ensuremath{\Delta}}   
\newcommand{\delayBound}{\ensuremath{\Delta^{ub}}}   
\newcommand{\transmissionRate}{\ensuremath{\alpha}}   
\newcommand{\rateBackhaul}{\ensuremath{\sigma}}   
\newcommand{\bandwidth}{\ensuremath{B}}   
\newcommand{\transmitPower}{\ensuremath{\rho}}   
\newcommand{\dist}{\ensuremath{\delta}}   
\newcommand{\channelFading}{\ensuremath{\hbar}}   
\newcommand{\whiteGaussian}{\ensuremath{w_0}}   
\newcommand{\pathLoss}{\ensuremath{\upsilon}}   
\newcommand{\offloadTime}{\ensuremath{t^{off}}}   
\newcommand{\downloadTime}{\ensuremath{t^{res}}}   
\newcommand{\serverProcTime}{\ensuremath{t^{proc}}}   
\newcommand{\prior}{\ensuremath{\pi}}   
\newcommand{\priorCand}{\ensuremath{\pi^{cand}}}   
\newcommand{\lastAssignedDeadline}{\omega}
\newcommand{\mapping}{\ensuremath{\mathcal{M}}} 
\newcommand{\entryTime}{\ensuremath{t^{entry}}} 
\newcommand{\exitTime}{\ensuremath{t^{exit}}} 
\newcommand{\stageUtil}{\ensuremath{u^{stg}}}
\newcommand{\deadlineRatio}{\ensuremath{\beta}}

\newcommand{\vehicleSet}{\ensuremath{\mathcal{V}}} 
\newcommand{\jobSet}{\ensuremath{\mathcal{J}}} 
\newcommand{\rsuSet}{\ensuremath{\mathcal{R}}} 
\newcommand{\serverSet}{\ensuremath{\mathcal{S}}} 
\newcommand{\feasibleRSUSet}{\ensuremath{\mathcal{Q}_\jobInd}} 
\newcommand{\tempSet}{\ensuremath{\mathcal{Q}}} 
\newcommand{\localJobsSet}{\ensuremath{\mathcal{C}}} 

\newcommand{\localVar}{\ensuremath{L}} 
\newcommand{\uploadVar}{\ensuremath{X}} 
\newcommand{\downloadVar}{\ensuremath{Z}} 
\newcommand{\procVar}{\ensuremath{Y}} 

\newcommand{\hpJobs}{\ensuremath{\mathcal{H}^{job}}}   
\newcommand{\hpiJobs}{\ensuremath{\mathcal{H}^{+}}}   
\newcommand{\lpJobs}{\ensuremath{\mathcal{L}^{job}}}   
\newcommand{\stg}{\ensuremath{S}}   
\newcommand{\procTime}{\ensuremath{P}}   
\newcommand{\effProcTime}{\ensuremath{t}}   
\newcommand{\indicator}{\ensuremath{\mathcal{I}}}   
\newcommand{\jobTermSeg}{\ensuremath{\xi}}  
\newcommand{\false}{{\fontfamily{lmtt}\selectfont False}}
\newcommand{\true}{{\fontfamily{lmtt}\selectfont True}}  
\newcommand{\brk}{{\fontfamily{lmtt}\selectfont Break}} 
\newcommand{\failure}{{\fontfamily{lmtt}\selectfont Failure} }
\newcommand{\return}{{\fontfamily{lmtt}\selectfont Return}} 
\newcommand{\continue}{{\fontfamily{lmtt}\selectfont Continue}} 

\newcommand{\problem}{{{\fontfamily{lmtt}\selectfont MMSV}}} 
\newcommand{\heur}{{{\fontfamily{lmtt}\selectfont MJMP}}} 
\newcommand{\baselineOne}{{{\fontfamily{lmtt}\selectfont EFF}}} 
\newcommand{\baselineDecomposition}{{{\fontfamily{lmtt}\selectfont DCMP}}}


\section{Introduction} \label{sec:introduction}

Vehicular Edge Computing (VEC) extends edge computing to intelligent transportation systems by enabling vehicles to offload computation-intensive applications to edge servers~\cite{liu2021vehicular, meneguette2021vehicular, gao2024energy}. It has emerged as a promising paradigm for supporting low-latency, computation-intensive applications. Since onboard computational resources are limited, jobs may be processed locally or offloaded to edge infrastructure.

Effective mapping of jobs to VEC resources and scheduling their access to these resources are crucial for fully exploiting the benefits of vehicular edge computing. In addition to resource constraints, vehicle \textit{mobility} introduces time-varying resource availability. Scheduling decisions must therefore \textit{jointly} decide whether to execute locally or offload a job, resource mapping, and execution order, while explicitly considering vehicle mobility and end-to-end deadlines.

Existing approaches often consider resource contention in isolation: some works ignore contention altogether~\cite{ma2024latency, zhang2024computational}, while others consider only computation~\cite{xu2023adaptive, gao2023joint, abdi2024task, sang2024mobility, tiwari2024rate} or communication resources~\cite{huang2023joint, huang2024mobility}. A smaller body of work considers contention across both communication and computation resources~\cite{pradhan2024towards, chauhan2024probabilistic, li2024uav, gao2024energy}. However, their models are restricted to two resource types, limiting their ability to capture more general and heterogeneous VEC systems.

Furthermore, existing approaches often rely on ILP and other computationally expensive optimization techniques~\cite{alameddine2019dynamic, lai2024short}. Such approaches are less suitable for VEC environments, where job arrivals are determined at runtime and resource availability can change dynamically, requiring scheduling decisions to be made with low overhead. These limitations motivate a lightweight, holistic resource mapping and scheduling approach that supports diverse resource types and configurations while preserving the end-to-end delay structure.

Emerging VEC applications may involve multiple stages spanning wireless and wired networks as well as heterogeneous computing resources, such as CPUs, GPUs, and other accelerators. Existing works also rarely consider the download of results back to the vehicle, despite it being part of the end-to-end processing path. Furthermore, different applications may follow different resource access patterns. This motivates modeling a VEC application as an \textit{end-to-end multi-stage pipeline}, where each stage represents communication or computation over a potentially heterogeneous resource. The resulting model provides a generic framework that can accommodate diverse resource configurations and access patterns without being tied to a specific system configuration. Furthermore, this model also support \textbf{both} preemptive and non-preemptive stages seamlessly.

Mobility has also been considered in recent VEC scheduling and resource allocation studies~\cite{zhu2018task, sorkhoh2019workload, sang2024mobility, huang2024mobility}. Our goal, however, is to develop a \textit{generic framework} for VEC applications. The model should capture arbitrary numbers and types of heterogeneous resources, resource contention across different stages, end-to-end timing constraints, and mobility, while supporting joint resource mapping and scheduling.

Specifically, we ask the following fundamental research question: \textit{How can multi-stage real-time scheduling and resource allocation approaches be adapted to Vehicular Edge Computing systems to jointly optimize resource mapping and task offloading under mobility and end-to-end timing constraints?}

\textbf{From Prior Work to Our Design}:
Recently, \cite{kumar2024optimal} studied fixed-priority (FP) scheduling of real-time jobs with end-to-end deadlines in distributed systems, focusing on multi-stage pipelines where each stage corresponds to a resource type and may consists of multiple heterogeneous resources. Their approach combines \textit{optimal priority assignment} (OPA)~\cite{audsley1991optimal} with a schedulability test based on \textit{delay-composition algebra} (DCA)~\cite{jayachandran2008delay, jayachandran2008transforming} to compute an \textbf{optimal} schedule. While the work also evaluates a VEC setting, it assumes that the mapping of jobs to resources is given. Building on this framework, we address the joint resource mapping and scheduling problem while explicitly incorporating vehicle mobility into these decisions.

Prior work~\cite{palencia2003offset, hong2011meeting} addresses real-time scheduling with end-to-end deadline constraints by \textit{decomposing} an end-to-end deadline into per-stage deadlines and then applying single-stage scheduling techniques. While these techniques can be applied in a VEC setting too, however, deadline decomposition is non-trivial and may turn an otherwise schedulable workload into an unschedulable one~\cite{kumar2024optimal}.

The key contributions of this work are summarized in the following.
\begin{itemize}
    \item In line with~\cite{kumar2024optimal}, we model a VEC system as a multi-stage multi-resource distributed real-time system, enabling the use of established results from this area. Building on~\cite{kumar2024optimal}, we additionally address the mapping of jobs to heterogeneous resources and explicitly account for vehicle mobility.

    \item We propose a mobility-aware heuristic \heur\ that jointly performs mapping and scheduling by utilizing optimal priority assignment and delay-composition algebra while accounting for mobility and end-to-end delay to minimize energy consumption.

    \item We conduct an extensive evaluation across a diverse range of system and workload configurations against two baselines. The first uses the same delay-composition algebra, isolating the impact of our mapping strategy, while the second uses a decomposition-based approach, highlighting the benefits of directly optimizing end-to-end delay.
\end{itemize}

\textbf{Extensions to \cite{kumar2024optimal}:} This work builds on the multi-stage multi-resource distributed scheduling framework developed in \cite{kumar2024optimal}, which proposes an optimal scheduling approach and uses a VEC setting for evaluation. However, \cite{kumar2024optimal} does not explicitly model vehicle mobility and assumes that resource mapping is given. In contrast, this work jointly addresses resource mapping and scheduling while explicitly accounting for mobility and its impact on resource availability. Furthermore, unlike \cite{kumar2024optimal}, we also consider the energy consumption of vehicles as an optimization objective.


\section{System Model and Problem Statement} \label{sec:model}

\begin{figure}
    \centering
    \includegraphics[width=0.8\linewidth]{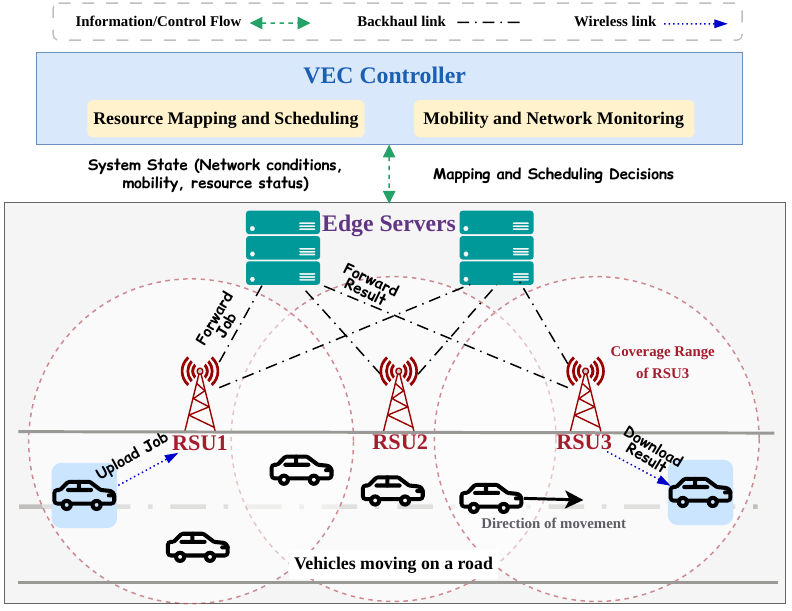}
    \caption{Vehicular Edge Computing System Architecture}
    \label{fig:system}
\end{figure}

\subsection{System Architecture}

Figure~\ref{fig:system} illustrates the Vehicular Edge Computing (VEC) architecture considered in this work. The system consists of a set of vehicles $\vehicleSet = \{\vehicle_\vehicleInd\}$, roadside units (RSUs) $\rsuSet = \{\rsu_\rsuInd\}$, and servers $\serverSet = \{\server_\serverInd\}$. Vehicles communicate with RSUs over a wireless network, while RSUs connect to servers through a wired backhaul network. We assume that vehicles move in the same direction. For simplicity, we assume that each RSU is connected to every server. More generally, RSU$\leftrightarrow$server connectivity can be represented by a connectivity set $\mathcal{C} \subseteq \rsuSet \times \serverSet$, where $(\rsu_i,\server_j) \in \mathcal{C}$ indicates that RSU $\rsu_i$ can communicate with server $\server_j$. The proposed approach does not require full connectivity and can directly accommodate arbitrary RSU$\leftrightarrow$server connectivity.

Following common practice in the literature, we assume that the wireless network supports simultaneous uploading and downloading for vehicles within the coverage range of an RSU, with uplink and downlink transmissions using orthogonal channels~\cite{gao2024energy}. Since vehicles are mobile, the RSU used for downloading may differ from the one used for uploading. 
Each server $\server_\serverInd$ provides a processing capacity of $\speedServer_\serverInd$ cycles per unit time. Vehicle mobility and network conditions are monitored by the \textit{control plane}, and the resulting information is made available to the VEC controller for resource mapping and scheduling decisions.

\subsection{Job Model}

We assume that the offloading algorithm runs periodically on the VEC controller, with each invocation receiving a \textit{snapshot} of the system representing its current state. For simplicity, each vehicle $\vehicle_\vehicleInd$ generates one job $\job_\jobInd$. Let $\jobSet$ denote the set of $\noJobs$ jobs. Each job $\job_\jobInd \in \jobSet$ is characterized by $\langle \position_\jobInd, \arrivalTime_\jobInd, \dataSizeInput_\jobInd, \dataSizeResult_\jobInd, \cyclesRequired_\jobInd, \deadline_\jobInd, \energyLocal_\jobInd \rangle$ as explained in the following.

Let $\position_\jobInd$ be the current absolute position of the corresponding vehicle $\vehicle_\vehicleInd$, $\arrivalTime_\jobInd \geq 0$ is its arrival time, $\dataSizeInput_\jobInd$ and $\dataSizeResult_\jobInd$ are the input and result sizes (in MB), respectively, and $\deadline_\jobInd$ is its deadline. Jobs may arrive in the future, for example, due to periodic tasks running on vehicles or event-triggered applications.

Given the vehicle speed $\speedVehicle_\vehicleInd$, we can determine the time window $[\rangeEnterTime_{\vehicleInd,\rsuInd},\rangeExitTime_{\vehicleInd,\rsuInd}]$ during which vehicle $\vehicle_\vehicleInd$ remains within the coverage range of RSU $\rsu_\rsuInd$. Here, $\rangeEnterTime_{\vehicleInd,\rsuInd}$ and $\rangeExitTime_{\vehicleInd,\rsuInd}$ denote the entry and exit times, respectively. The corresponding \textit{sojourn time} is
\begin{equation}
    \sojournTime_{\vehicleInd,\rsuInd} = \rangeExitTime_{\vehicleInd,\rsuInd} - \rangeEnterTime_{\vehicleInd,\rsuInd}.
\end{equation}
Thus, any upload or download through an RSU must be completed within the corresponding coverage window. Moreover, let $\cyclesRequired_\jobInd$ denotes the number of CPU cycles required to complete $\job_\jobInd$.

A job can either be executed locally or offloaded to a server through an RSU. For local execution, we assume that the onboard processor can complete the job before its deadline at an energy cost of $\energyLocal_\jobInd$. For offloaded execution, the vehicle consumes energy for uploading the job and downloading the result. For simplicity, we assume that every server provides the resource types required to process any job, allowing a job to be assigned to any server. This assumption can be relaxed trivially by restricting the set of feasible mappings.

\subsection{Delay and Energy Model}

Following common practice, the transmission rate between vehicle $\vehicle_\vehicleInd$ and RSU $\rsu_\rsuInd$ is given by
\begin{equation}
    \transmissionRate_{\jobInd,\rsuInd} = \bandwidth \log_2 \left( 1 + \frac{\transmitPower_\jobInd |\channelFading|^2}{\whiteGaussian\left(\dist_{\jobInd,\rsuInd}\right)^{\pathLoss}} \right),
\end{equation}
where $\bandwidth$ is the bandwidth, $\transmitPower_\jobInd$ is the transmit power of $\vehicle_\vehicleInd$, $\channelFading$ is the channel fading coefficient, $\whiteGaussian$ is the white Gaussian noise power, $\dist_{\jobInd,\rsuInd}$ is the distance between vehicle $\vehicle_\vehicleInd$ and RSU $\rsu_\rsuInd$, and $\pathLoss$ is the path-loss exponent.
Thus, the time required to upload the job $\job_\jobInd$ to RSU $\rsu_\rsuInd$ is
\begin{equation} \label{eq:offloading_time}
\offloadTime_{\jobInd,\rsuInd} = \frac{\dataSizeInput_\jobInd}{\transmissionRate_{\jobInd,\rsuInd}}
\end{equation}
while downloading the result takes
\begin{equation} \label{eq:downloading_time}
\downloadTime_{\jobInd,\rsuInd} = \frac{\dataSizeResult_\jobInd}{\transmissionRate_{\jobInd,\rsuInd}}
\end{equation}

We assume a high-speed backhaul network which provides a constant transmission rate $\rateBackhaul$, resulting in an input transfer delay of $\frac{\dataSizeInput_\jobInd}{\rateBackhaul}$ from the RSU to the server and a result transfer delay of $\frac{\dataSizeResult_\jobInd}{\rateBackhaul}$ from the server to the RSU.

The processing time of job $\job_\jobInd$ on server $\server_\serverInd$ is
\begin{equation} \label{eq:proc_time}
\serverProcTime_{\jobInd,\serverInd} = \frac{\cyclesRequired_\jobInd}{\speedServer_\serverInd}
\end{equation}

We allow preemptive execution at servers, while wireless as well as backhaul transmissions are non-preemptive. The end-to-end delay $\delay_\jobInd$ therefore depends not only on its upload, processing, and download times, but also on interference from higher-priority jobs at preemptive stages and from both higher- and lower-priority jobs at non-preemptive stages.

The energy consumed by vehicle $\vehicle_\vehicleInd$ for local execution of job $\job_\jobInd$ is $\energyLocal_\jobInd$. If the job is offloaded through RSU $\rsu_\rsuInd$ and its result is downloaded through RSU $\rsu_\tempInd$, the corresponding energy consumption is $\transmitPower_\jobInd\left(\offloadTime_{\jobInd,\rsuInd}+\downloadTime_{\jobInd,\tempInd}\right)$.

Let $\uploadVar_{\jobInd,\rsuInd}$ be a binary variable that is 1 if job $\job_\jobInd$ is uploaded through RSU $\rsu_\rsuInd$, and 0 otherwise. Similarly, $\downloadVar_{\jobInd,\rsuInd}$ indicates whether the result is downloaded through RSU $\rsu_\rsuInd$, and $\procVar_{\jobInd,\serverInd}$ indicates whether the job is processed on server $\server_\serverInd$. Let $\localVar_\jobInd$ indicate whether the job is executed locally ($\localVar_\jobInd=1$) or remotely ($\localVar_\jobInd=0$). The total vehicle energy consumption is then
\begin{equation} \label{eq:total_energy}
\energyTotal =
\sum_{\jobInd=1}^{\noJobs}
\left[
\localVar_\jobInd \  \energyLocal_\jobInd
+
\transmitPower_\jobInd
\sum_{\rsuInd}
\left(
\uploadVar_{\jobInd,\rsuInd} \ \offloadTime_{\jobInd,\rsuInd}
+
\downloadVar_{\jobInd,\rsuInd} \ \downloadTime_{\jobInd,\rsuInd}
\right)
\right]
\end{equation}

For a locally executed job, $\localVar_\jobInd=1$, and all upload, processing, and download variables are zero. For an offloaded job, $\localVar_\jobInd=0$, with exactly one upload RSU, one processing server, and one download RSU selected.

Let $\mapping_{\jobInd,\stgInd}$ denote the resource assigned to job $\job_\jobInd$ in stage $\stg_\stgInd$. For example,
$\uploadVar_{\jobInd,\rsuInd}=1$ if $\mapping_{\jobInd,\stgInd}=\rsu_\rsuInd$ for the corresponding upload stage. 

\subsection{Problem Formulation}\label{subsec:formulation}

Given the vehicles $\vehicleSet=\{\vehicle_\vehicleInd\}$, RSUs $\rsuSet=\{\rsu_\rsuInd\}$, servers $\serverSet=\{\server_\serverInd\}$, and available jobs $\jobSet=\{\job_\jobInd\}$, the \underline{m}obility aware \underline{m}apping and \underline{s}cheduling in \underline{V}EC problem, henceforth referred to as \problem, is to determine, for each job $\job_\jobInd$:
\begin{enumerate}
\item whether the job is executed locally or offloaded, as indicated by $\localVar_\jobInd$;
\item if offloaded, the RSU used for uploading, the server used for processing, and the RSU used for downloading, represented by $\uploadVar_{\jobInd,\rsuInd}$, $\procVar_{\jobInd,\serverInd}$, and $\downloadVar_{\jobInd,\rsuInd}$, respectively;
\item a unique global priority $\prior_\jobInd$ for each job, which determines its execution order across the pipeline. The resulting schedule must account for the set of higher-priority jobs at preemptive stages and both higher- and lower-priority jobs at non-preemptive stages.
\end{enumerate}
The resulting schedule must ensure that every offloaded job completes within its end-to-end deadline:
\begin{equation}
    (1-\localVar_\jobInd)\delay_\jobInd \leq \deadline_\jobInd
\end{equation}
where $\delay_\jobInd$ is the end-to-end deadline of $\job_\jobInd$ in the pipeline with both preemptive and non-preemptive stages. The objective is to minimize the total energy consumption of the vehicles $\energyTotal$ in Eq.~\eqref{eq:total_energy}.

\textbf{Problem Complexity}: Even a restricted version of \problem\ reduces to the \textit{flow-shop} scheduling problem with job deadlines, which is NP-hard~\cite{bettati1992end, gao2024energy}. Specifically, by fixing vehicle mobility, resource mappings, and energy considerations, and assuming a fixed sequence of resource stages, \problem\ reduces to scheduling jobs with end-to-end deadlines on a flow-shop system. Therefore, \problem\ is NP-hard. The full problem is further complicated by vehicle mobility and the resulting dynamic resource availability, which must be jointly considered with resource mapping, end-to-end deadline constraints, and priority assignment. Moreover, the scheduler must optimize vehicle energy consumption while maintaining deadline feasibility.


\section{Proposed Approach} \label{sec:proposed}
In the following, we present our \underline{M}obility-Aware \underline{J}oint \underline{M}apping and \underline{P}riority Assignment (\heur) approach for solving \problem. The problem can be viewed as three closely related subproblems: \texttt{[SP1]}, end-to-end delay computation; \texttt{[SP2]}, priority assignment; and \texttt{[SP3]}, mobility-aware resource mapping for each stage. We address \texttt{SP1} using the Delay-Composition Algebra (DCA) described in Section~\ref{sec:dca} and \texttt{SP2} using the Optimal Priority Assignment (OPA) described in Section~\ref{sec:opa}. Finally, since \problem\ is NP-hard, we develop an efficient mobility-aware heuristic, \heur, to jointly address resource mapping and priority assignment, as described in Section~\ref{sec:heur}.
Notably, the formulation in Section~\ref{sec:model} can be extended to an ILP; however, such a solution has limited utility due to scalability and the low runtime requirement in a VEC setting.

\subsection{Optimal Priority Assignment (OPA)} \label{sec:opa}

OPA assigns \textit{fixed priorities} to a set of $\noTasks$ tasks by constructing a \textit{linear ordering} (permutation) of the tasks in $O(\noTasks^2)$ time~\cite{audsley1991optimal}. The priorities are assigned incrementally, starting from the lowest and proceeding toward the highest. At each priority level, OPA considers the tasks that have not yet been assigned a priority and identifies one that can feasibly take the current priority. Feasibility is checked using an \textit{OPA-compatible} feasibility test~\cite{davis2009priority, davis2016review, kumar2024optimal}. Once a feasible task is found, OPA assigns it the current priority and removes it from further consideration. If no remaining task passes the feasibility test, OPA declares the task set unschedulable. The procedure continues until all tasks have been assigned priorities.

As shown in~\cite{kumar2024optimal}, an OPA-compatible feasibility test can be constructed using DCA. In our setting, OPA assigns each offloaded job $\job_\jobInd$ a unique global fixed priority $\prior_\jobInd$. These priorities define the ordering of jobs when accessing shared resources throughout the processing pipeline.
In Section~\ref{sec:heur}, we use OPA for joint priority assignment and job-to-resource mapping.

\subsection{Delay Composition Algebra (DCA)} \label{sec:dca}

DCA provides an \textit{upper bound} on the end-to-end delay of a job executing in a multi-stage pipeline by accounting for the overlap in job execution across stages. We briefly describe the delay-bound computation below to make the paper self-contained; additional details can be found in~\cite{jayachandran2008delay, jayachandran2008transforming, kumar2024optimal}.

\begin{figure}
    \centering
    \includegraphics[width=0.9\linewidth]{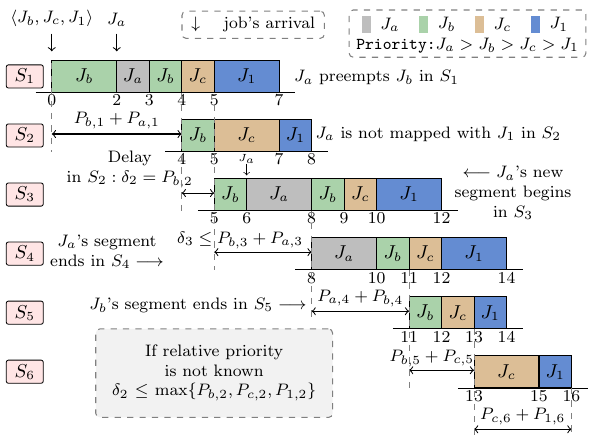}
    \caption{End-to-End Delay in a Preemptive Pipeline}
    \label{fig:dca}
\end{figure}

Let $\job_1$ be the target job for which we want to compute a delay bound $\delayBound_1$ such that $\delayBound_1 \geq \delay_1$. The job executes alongside a set of higher-priority jobs $\hpJobs_1$ and lower-priority jobs $\lpJobs_1$ in a pipeline with $\noStg$ stages, as illustrated in Figure~\ref{fig:dca}. We exclude from $\hpJobs_1$ and $\lpJobs_1$ any jobs whose activation windows do not overlap with that of $\job_1$. We further define $\hpiJobs_1 = \hpJobs_1 \cup {\job_1}$. Let $\procTime_{\jobInd,\stgInd}$ denote the processing time of job $\job_\jobInd$ at stage $\stg_\stgInd$.

As illustrated in Figure~\ref{fig:dca}, if the relative priority among jobs and their arrival times at the first stage are known, the \textit{exact} end-to-end delay can be computed incrementally. Specifically, the arrival and finish times of jobs are determined at each stage, traversing the pipeline from the first to the last stage and considering jobs in decreasing priority order. However, OPA compatibility requires only the sets of higher- and lower-priority jobs to be known; their relative ordering within each set is not yet determined. DCA avoids this requirement by deriving a delay bound based on the overlap of job execution across stages.

The DCA bound consists of a \textit{stage-additive} component for each stage except the last and a \textit{job-additive} component for higher-priority jobs. To formulate these components, we define a \textit{segment} with respect to job $\job_\jobInd$ as a sequence of consecutive stages such that (i) $\job_\jobInd$ shares every stage in the segment with $\job_1$, and (ii) the first (last) stage of the segment is either the first (last) stage of the pipeline or $\job_\jobInd$ does not share the stage immediately before (after) the segment with $\job_1$. For example, $\langle\stg_3,\stg_4\rangle$ forms a two-stage segment of job $\job_a$. Let $\noSeg_\jobInd$ denote the number of segments of job $\job_\jobInd$.

The end-to-end delay bound is composed of the following terms: (1). for each stage except the last, the maximum processing time among jobs in $\hpiJobs_1$ that share the stage with $\job_1$;
(2). for each one-stage segment, the processing time of the corresponding higher-priority job;
(3). for each segment spanning two or more stages, the highest and second-highest processing times of the corresponding higher-priority jobs within the segment under preemptive scheduling, and only the highest processing time under non-preemptive scheduling; and
(4). under non-preemptive scheduling, a blocking term for each stage, given by the maximum processing time among all jobs that share that stage.
These rules summarize the DCA formulation presented in~\cite{jayachandran2008delay, jayachandran2008transforming}.
We define the \emph{effective} processing time of job $\job_\jobInd$ at stage $\stg_\stgInd$ as
$$
\effProcTime_{\jobInd,\stgInd} =
\begin{cases}
\procTime_{\jobInd,\stgInd}, & \text{if $\job_\jobInd$ and $\job_1$ share stage $\stg_\stgInd$},\\
0, & \text{otherwise}.
\end{cases}
$$
and an indicator variable $\indicator_\stgInd$ which is 1 for a non-preemptive stage and 0 for a preemptive stage. Using these definitions, the end-to-end delay bound $\delayBound_1$ for a pipeline containing both preemptive and non-preemptive stages is given below.
\begin{equation} \label{eq:delay_bound_dca}
\delayBound_1 =
\sum\limits_{\stgInd=1}^{\noStg-1}
\max_{\job_\jobInd \in \hpiJobs_1}
{\effProcTime_{\jobInd,\stgInd}}
+
\sum\limits_{\job_\jobInd \in \hpiJobs_1}
\jobTermSeg_{\jobInd}
+
\sum\limits_{\stgInd=1}^{\noStg}
\indicator_\stgInd
\max\limits_{\job_\jobInd \in \jobSet \setminus \job_1}
{\effProcTime_{\jobInd,\stgInd}}
\end{equation}
where $\jobTermSeg_{\jobInd}$ denotes the interference terms due to higher priority jobs computed as per the rules discussed above. 
Note that Eq.~\eqref{eq:delay_bound_dca} builds upon the delay bounds for preemptive and non-preemptive scheduling given in Eqs.~(5) and (6) of~\cite{kumar2024optimal}. Therefore, the OPA-compatibility proofs established in Section IV.A~\cite{kumar2024optimal} can likewise be extended to Eq.~\eqref{eq:delay_bound_dca}.

\textbf{Adapting DCA to Our VEC Setting.} The generic delay bound in Eq.~\eqref{eq:delay_bound_dca} consists of three components: a stage-additive component capturing the execution time across stages, a job-additive component capturing interference from higher-priority jobs, and a blocking component capturing interference from lower-priority jobs at non-preemptive stages. Our VEC model consists of five stages. Each processing term is accounted exactly once, either as a stage-additive term, a job-additive term due to higher-priority jobs, or a job-blocking term due to a non-preemptive stage.

\subsection{Mobility aware Joint Mapping and Priority Assignment} \label{sec:heur}

Since mapping is NP-hard, combinatorial approaches such as ILP, despite providing optimal solutions, are not well suited to our online setting, where the complete set of jobs is not known in advance. We therefore propose a lightweight mapping and priority-assignment approach, \heur, that can be applied online to obtain a near-optimal solution. \heur\ jointly determines resource mappings and global priorities while explicitly accounting for vehicle mobility to optimize the energy consumption of vehicles.

The key insight behind \heur\ is to decouple mobility from the mapping of intermediate stages. Specifically, processing at the intermediate stages is independent of vehicle mobility, which affects only the first and last stages of the pipeline. A job can be uploaded to an RSU, or its result downloaded from an RSU, only while the vehicle is within the corresponding coverage range. The fundamental idea behind the mapping strategy is to distribute jobs across available resources so that jobs with similar deadlines are not concentrated on the same resource. This helps reduce contention and improves the likelihood of meeting end-to-end deadlines, while RSU mappings additionally account for the mobility and sojourn time. Priority assignment follows OPA. The overall procedure is summarized in Algorithm~\ref{alg:heur} and discussed below.

\begin{algorithm}[t]
\caption{\heur\ -- Mobility aware Joint Mapping and Priority Assignment} \label{alg:heur}

    $\localJobsSet \gets \emptyset$; $\priorCand \gets \noJobs$ \label{step:algo_heur_initialization1} \;
    
    $\mapping^{RU} \gets \textsc{RSUUploadMapping()}$\;
    $\mapping^{S} \gets \textsc{ServerMapping()}$\;
    
    Sort $\jobSet$ in the increasing order of energy density\;
    $\feasibleRSUSet \gets \textsc{GetCandRSUDownload()}, \ \forall \jobInd$ \label{step:algo_heur_feasible_rsu_set}  \; 
    \ForEach{$\job_\jobInd \in \jobSet \setminus \localJobsSet $}{
        $\textsc{Rearrange(\feasibleRSUSet)}$ \label{step:algo_heur_rsu_right_to_left}\;
        $status \gets$ \false\;
        \ForEach{$\rsu_\rsuInd \in \feasibleRSUSet$}{
            \If{$\rsu_\rsuInd$ is feasible with $\priorCand$}{
                $\prior_\jobInd \gets \priorCand$; $\mapping_{\jobInd, 5} \gets \rsu_\rsuInd$ \; 
                $\priorCand \gets \priorCand - 1$; $status \gets$ \true; \brk\;
            }
        }
        \If{$!status$}{
            $\localJobsSet \gets \localJobsSet  \cup \job_\jobInd$; Update $\mapping^{RU}$ and $\mapping^{S}$ \label{step:algo_heur_migrate_local} \;
        }
    }
    $\mapping^{RD} \gets \{\mapping_{\jobInd, 5}\}$\; 
    $\mapping^{RS} \gets \textsc{RSUServerLinkMapping}\left({\mapping^{RU}, \mapping^{S}}\right)$\; 
    $\mapping^{SR} \gets \textsc{ServerRSULinkMapping}\left(\mapping^{S}, \mapping^{RD}\right)$\; 
    
    \Return{$\langle \{\pi\},\mapping,\localJobsSet \rangle$}\;

\end{algorithm}

\heur\ first computes the upload mapping $\mapping^{RU}={\mapping_{\jobInd,1}}$ using $\textsc{RSUUploadMapping()}$, given in Algorithm~\ref{alg:rsu_upload_mapping}. Specifically, $\textsc{RSUUploadMapping}()$ first orders jobs by their absolute deadlines, $\arrivalTime_\jobInd + \deadline_\jobInd$. For each job, it selects the RSU with the smallest deadline assigned so far among those for which the vehicle's sojourn time is sufficient to complete the upload (Step~\ref{step:algo_rsu_upload_select_rsu}). Thus, an RSU is excluded if the vehicle has already left its coverage range or if the remaining sojourn time is insufficient for the upload. Since uploading is non-preemptive, once an upload begins, it must complete before another upload can use the same transmission resource. Thus, if no feasible RSU is found, then in Step~\ref{step:algo_rsu_mapping_migrate_local} the job is migrated for local execution.

The server mapping is then determined in Algorithm~\ref{alg:heur} using $\textsc{ServerMapping()}$, which follows the same basic principle of spreading jobs across resources according to their deadlines. Unlike the upload stage, server mapping is not directly constrained by mobility; under our assumption that every server can process every job, all servers are feasible and no additional feasibility test is required.

Next, the set of locally executed jobs $\localJobsSet$ is initialized to $\emptyset$, and the candidate priority $\priorCand$ used by OPA is initialized to $\noJobs$ in Step~\ref{step:algo_heur_initialization1}. Jobs are then considered in increasing order of \textit{energy density}, defined as energy consumption per unit of processing time. This ordering allows the heuristic to favor local execution for jobs that are relatively expensive to execute locally when selecting between local and offloaded execution.

\begin{algorithm}[t]
\caption{$ \textsc{RSUUploadMapping()}$} \label{alg:rsu_upload_mapping}
    
    Sort $\jobSet = \{\job_\jobInd\}$ in increasing order of absolute deadline $\arrivalTime_{\jobInd}+\deadline_{\jobInd}$\;
    $\lastAssignedDeadline_{\rsuInd} \gets 0, \forall \rsuInd | \rsu_\rsuInd \in \rsuSet$\;
    \ForEach{$\job_\jobInd \in \jobSet$}{
        $\feasibleRSUSet \gets \emptyset$\;
        \ForEach{$\rsu_\rsuInd \in \rsuSet$}{
            \If{$\arrivalTime_{\jobInd} \leq \exitTime_{\jobInd,\rsuInd} - \offloadTime_{\jobInd,\rsuInd}$}{
                $\feasibleRSUSet \gets \feasibleRSUSet \cup \rsu_\rsuInd$\;
            }
        }
        \If{$\feasibleRSUSet \neq \emptyset$}{
            $\mapping_{\jobInd, 1} \gets \arg\min\limits_{\rsu_\rsuInd \in \feasibleRSUSet} \lastAssignedDeadline_{\rsuInd}$; \label{step:algo_rsu_upload_select_rsu}
            $\lastAssignedDeadline_{\rsuInd} \gets \arrivalTime_{\jobInd}+\deadline_{\jobInd}$\;
        }
        \Else{
            $\localJobsSet \gets \localJobsSet  \cup \job_\jobInd$\label{step:algo_rsu_mapping_migrate_local} \;
        }
    }
\Return{$\{\mapping_{\jobInd, 1}\}$}\;
\end{algorithm}

The priority assignment and download mapping are determined in Steps~\ref{step:algo_heur_rsu_right_to_left}--\ref{step:algo_heur_migrate_local}. For each job, a set of feasible RSUs is first computed in Step~\ref{step:algo_heur_feasible_rsu_set}, initially assuming no interference from other jobs. The feasible set $\feasibleRSUSet$ is then reordered using $\textsc{Rearrange}(\feasibleRSUSet)$ so that RSUs are considered from right to left according to their absolute positions. This ordering allows the heuristic to select an RSU that is feasible under the current candidate priority $\priorCand$ while delaying the job as much as possible. If no feasible combination of priority and download RSU is found, the job is migrated to local execution.

The backhaul mappings are straightforward under the assumed full RSU$\leftrightarrow$server connectivity. Given the upload and server mappings $\langle\mapping^{RU},\mapping^{S}\rangle$, the RSU-to-server mapping $\mapping^{RS}$ can be directly determined. Similarly, given $\langle\mapping^{S},\mapping^{RD}\rangle$, the server-to-RSU mapping $\mapping^{SR}$ follows directly. No additional mapping decisions are therefore required for these stages.

Finally, \heur\ imposes a release restriction on the download stage. Although the execution of the intermediate stages is determined solely by priority, downloading cannot begin until the preceding stage has completed and the vehicle is within the coverage range of the selected RSU. Since the download mapping is determined using the \textbf{upper-bound delay}\footnote{The actual completion time of the previous stage may be earlier, allowing the download to begin before the delay bound.} provided by DCA, the download must begin at this delay bound to guarantee that the vehicle remains within range of the respective RSU. Thus, the download stage is governed by both global priority and release time, with the latter determined by the completion of the preceding stage.

\subsection{Discussion}

\textbf{Release Restrictions:}
The release restriction is introduced solely to ensure that the DCA-based upper-bound delay remains valid in the presence of mobility. It does not constitute an additional scheduling decision; rather, it constrains when the final transmission may start so that the delay bound remains consistent with the vehicle's movement. The restriction is needed only when mobility introduces uncertainty that must be captured by the upper-bound delay and therefore need not be enforced throughout the entire execution. 

\textbf{Why OPA?}
OPA can produce EDF- or RM-like priority orderings as special cases, but its main advantage in our distributed setting is its flexibility to assign priorities based on actual schedulability. Jobs need not be mapped to the same resources at every stage. For example, consider two jobs $J_1$ and $J_2$ with $D_1<D_2$. An EDF-like policy would \textit{always} assign $J_1$ higher priority than $J_2$ on every resource. However, given their different resource mappings, this ordering may make $J_2$ infeasible on one or more of its mapped resources, whereas reversing their priorities may allow both jobs to meet their deadlines. OPA therefore avoids imposing a globally deadline-ordered priority and instead determines priorities based on the schedulability of the mapped jobs. This flexibility is particularly important when jobs follow different resource paths through the system. Notably, for a given mapping and a schedulability test, OPA is optimal~\cite{davis2011survey, kumar2024optimal}.

\textbf{Why DCA?}
DCA is needed to compute delay bounds during the joint mapping and priority-assignment process. Under OPA, the complete priority ordering is not known until the priority-assignment procedure is completed. DCA provides the required delay bounds as priorities and mappings are progressively determined, allowing their feasibility to be evaluated during the procedure.


\section{Experimental Results} \label{sec:experiments}

We evaluate the performance of \heur\footnote{Upon acceptance, the source code will be made publicly available.} against two baselines across a wide range of system and workload configurations.
The first baseline \baselineOne\ uses the same DCA-based delay-computation framework as \heur, while the second \baselineDecomposition\ adopts a decomposition-based approach that assigns per-stage deadlines to satisfy an end-to-end deadline constraint.

We assume that the backhaul is a high-speed network, such as an optical-fiber network, with sufficient bandwidth to provide near-contention-free communication. Accordingly, we model the backhaul using a constant transmission rate and negligible queuing delay~\cite{alameddine2019dynamic, gao2024energy}. In our experiments, we therefore assume a fixed backhaul delay determined by the corresponding transmission rate. We also assume that all RSUs are available for offloading. The remaining system parameters are selected based on VEC configurations commonly used in the literature including~\cite{gao2024energy, kumar2024optimal, alameddine2019dynamic}.

\noindent
\textbf{Baselines:}
We consider two baselines. The first, \baselineOne, is an end-to-end-based configuration that uses EDF for priority assignment. Its mapping strategy operates as follows. For each RSU, jobs are ordered according to their arrival times. The first job in each RSU queue is considered in each iteration and assigned to the corresponding resource, after which the selected job is removed from all queues. A job that cannot satisfy its end-to-end deadline is migrated to local execution.

The second baseline, \baselineDecomposition, uses a decomposition-based approach. Each job is decomposed into a set of per-stage tasks with corresponding offsets and deadlines derived from its end-to-end deadline. The end-to-end deadline is distributed across the stages in proportion to their processing times, where the minimum processing time of each stage is used for the decomposition. \baselineDecomposition\ then selects the first available resource for each decomposed task. Feasibility is determined using the \textit{Demand Bound Function} (DBF). If a job is found to be infeasible, it is migrated to local execution and its corresponding decomposed tasks are removed from all stages.

\noindent
\textbf{Experimental Setup:}
We vary the number of jobs $\noJobs$, the number of resources in each stage $\noRes_\stgInd$, the average resource utilization in each stage $\stageUtil_\stgInd$, and $\beta$ which determines the deadline. Specifically, the deadline of each job is set to $\beta \times \xi$, where $\xi$ denotes the maximum time for which the corresponding vehicle remains on the road. 
Table~\ref{tab:simulation_parameters} shows the values of simulation parameters. Unless otherwise stated, the bold-faced values represent the default configuration.
$\beta$ is sampled uniformly from the specified range for each configuration in Table~\ref{tab:simulation_parameters}; for the default configuration, it is sampled uniformly from $[0.8,1]$.
For each configuration we generate 1000 test cases.

\begin{table}[t]
    \centering
    \caption{Simulation Parameter Configurations}
    \label{tab:simulation_parameters}
    \setlength{\tabcolsep}{3pt}
    \renewcommand{\arraystretch}{1.1}
    \resizebox{\columnwidth}{!}{%
    \begin{tabular}{lcccc}
        \toprule
        \textbf{Parameter}
        & \textbf{cfg1}
        & \textbf{cfg2}
        & \textbf{cfg3}
        & \textbf{cfg4} \\
        \midrule
        
        $\beta$
        & $[0.6,0.7]$
        & $[0.7,0.8]$
        & $[0.8,0.9]$
        & $[0.9,1]$ \\
        
        $\noRes_\stgInd$
        & $\{5,3,5\}$
        & $\mathbf{\{10,5,10\}}$
        & $\{15,7,15\}$
        & -- \\

        $\stageUtil_\stgInd$
        & $\mathbf{\{0.2,0.3,0.1\}}$
        & $\{0.2,0.4,0.1\}$
        & $\{0.2,0.5,0.1\}$
        & $\{0.3,0.6,0.1\}$ \\

        $\noJobs$
        & $80$
        & $\mathbf{100}$
        & $120$
        & -- \\

        \bottomrule
    \end{tabular}%
    }
\end{table}

We evaluate the approaches in terms of energy consumption at the vehicles. In particular, we focus on the energy saved through offloading \textit{relative} to executing \textit{all jobs locally}. Since the energy consumed for wireless transmission is negligible compared with the energy required for local processing in the considered setting, we ignore the energy cost of uploading jobs and downloading results in the experiments.

\noindent
\textbf{Results:}
For consistency across experiments, we report energy improvement relative to executing all jobs locally. Specifically, for each approach, the reported value is the saving in energy consumption compared to executing all jobs locally. Thus, a larger improvement indicates greater energy savings. 

\begin{figure}
    \centering
    \includegraphics[width=0.7\linewidth]{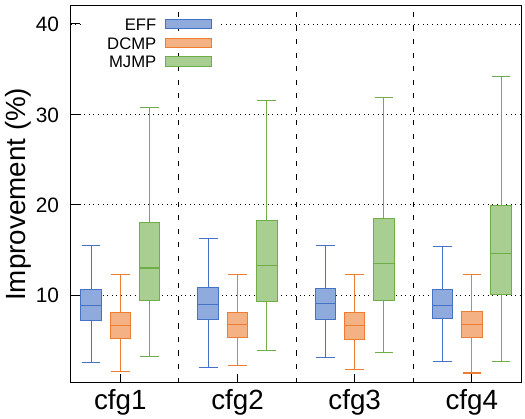}
    \caption{Varying Deadline}
    \label{fig:varying_deadline}
\end{figure}

Figure~\ref{fig:varying_deadline} shows the impact of the deadline ratio $\beta$ on energy savings. Since stage utilization remains constant, increasing $\beta$ primarily allows jobs to be assigned longer processing times, and therefore has only a limited impact on energy savings. This trend is reflected in our results. For $\beta\in[0.6,0.7]$, \baselineDecomposition, \baselineOne, and \heur\ achieve average improvements of 9\%, 6.7\%, and 14.1\%, respectively, relative to executing all jobs locally. For $\beta\in[0.9,1]$, the corresponding improvements are 9.8\%, 6.8\%, and 15.5\%. Thus, varying $\beta$ results in minor savings in average energy consumption. In contrast, varying stage utilization leads to a noticeable relative improvement for \heur, as discussed next. Overall, although the minimum energy savings of \heur\ are only slightly higher than those of the baselines or lower $\beta$ values, both its average and maximum savings consistently increase with $\beta$.

\begin{figure}
    \centering
    \includegraphics[width=0.7\linewidth]{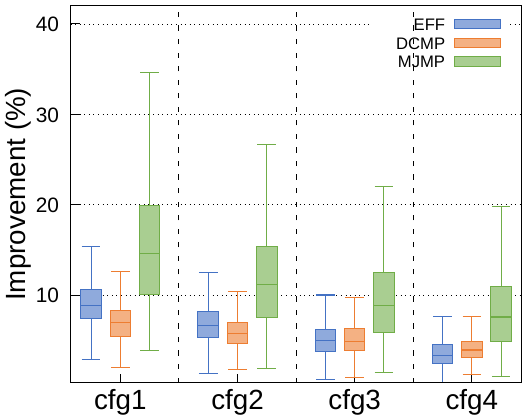}
    \caption{Varying Utilization}
    \label{fig:varying_util}
\end{figure}

Figure~\ref{fig:varying_util} shows the impact of resource utilization on energy savings. To isolate the effect of the first two stages, we keep the utilization of the final stage fixed while increasing the utilization of the first and second stages. Note, this assumptions reflect a realistic system. As utilization increases, jobs require a larger share of the available processing capacity, making fewer jobs feasible for offloading when the other parameters remain unchanged. Consequently, the energy savings generally decrease for all approaches. \heur, however, consistently maintains higher energy savings than both baselines. \baselineDecomposition\ performs slightly better than or comparably to \baselineOne\ across the considered configurations, but the gap between the baselines and \heur\ remains evident.

\begin{figure}
    \centering
    \includegraphics[width=0.6\linewidth]{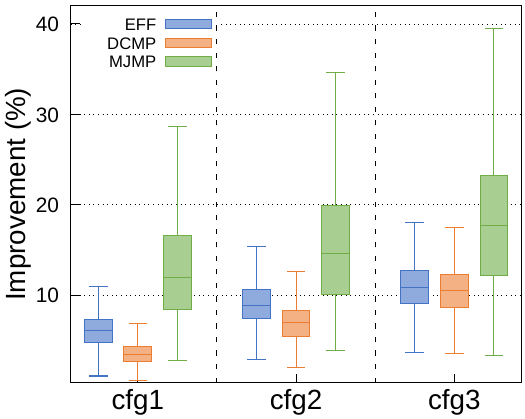}
    \caption{Varying Number of Resources}
    \label{fig:varying_resources}
\end{figure}

Figure~\ref{fig:varying_resources} illustrates the impact of the number of resources on energy savings. Increasing the number of resources provides greater flexibility for mapping jobs and reduces resource contention, allowing more jobs to be offloaded successfully. For example, when the number of resources increases from $\{5,3,5\}$ to $\{15,7,15\}$, the energy savings of \baselineDecomposition\ increase from 3.6\% to 10.4\% compared to executing all jobs locally. The performance of \baselineOne\ also increase from 6.1\% to 11\% compared to executing all jobs locally. In comparison, \heur\ improves considerably from 12.9\% to 18.2\% compared to executing all jobs locally. The improvement achieved by \heur\ reflects its ability to exploit the additional mapping flexibility when more resources are available. Moreover, the maximum achieved saving also increases as expected. Note that adding more resources also tends to increase the utilization assigned to each job, since the stage utilization and number of jobs remain constant. This explains the consistent but proportional improvement.

\begin{figure}
    \centering
    \includegraphics[width=0.6\linewidth]{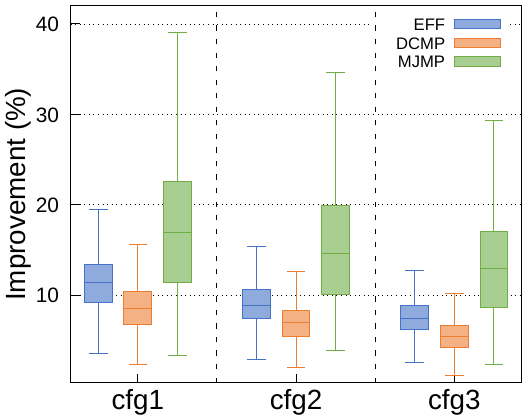}
    \caption{Varying Number of Jobs}
    \label{fig:varying_jobs}
\end{figure}

Figure~\ref{fig:varying_jobs} examines the effect of increasing the number of jobs while keeping the other parameters, including resource utilization, fixed. As the number of jobs increases, the same utilization is distributed among a larger number of jobs, resulting in smaller and more similar jobs. Although the fraction of jobs executed locally may remain relatively stable, the energy benefit of offloading each individual job decreases because the difference between local and remote execution becomes smaller. Consequently, the overall energy savings decrease as the number of jobs increases. The results show a consistent downward trend across the considered configurations, as expected.

\begin{figure}
    \centering
    \includegraphics[width=0.7\linewidth]{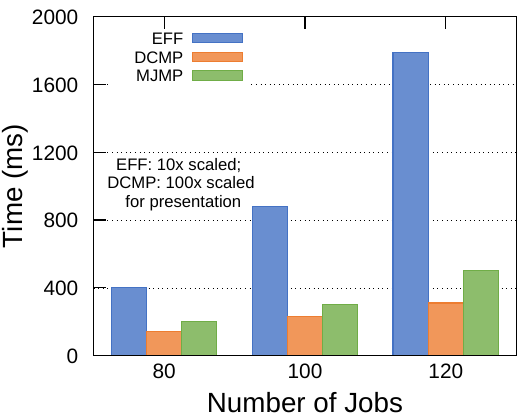}
    \caption{Runtime with Number of Jobs}
    \label{fig:runtime}
\end{figure}

Finally, Figure~\ref{fig:runtime} shows the average execution time of the algorithms as the number of jobs increases, providing an indication of their scalability. For presentation clarity, the runtime of \baselineDecomposition\ is scaled by $100\times$ and that of \baselineOne\ by $10\times$. As expected, the runtime increases with the number of jobs because the algorithms must process a larger set of jobs and perform more feasibility checks. Nevertheless, the runtime of \heur\ remains practical for the considered VEC configurations, demonstrating that the proposed approach can scale to large workloads while providing substantially higher energy savings.

\section{Conclusion} \label{sec:conclusion}

In this work, we studied joint resource mapping and scheduling for Vehicular Edge Computing (VEC) systems under vehicle mobility and end-to-end real-time constraints. We modeled VEC applications as multi-stage multi-resource distributed real-time systems. This formulation provides a flexible framework for representing a range of VEC applications with different resource configurations and access patterns.

Building on this model, we proposed \heur, a mobility-aware heuristic that jointly determines resource mappings and global job priorities. 
Mobility is incorporated through feasibility checks for the RSU mappings, allowing the approach to account for the limited time vehicles remain within RSU coverage. 
Extensive simulation across a wide range of system and workload configurations demonstrated the effectiveness of the proposed approach. \heur\ consistently achieved higher energy savings than both the DCA-based baseline and the decomposition-based approach. 

Overall, the results demonstrate that explicitly accounting for mobility, resource mapping, and end-to-end scheduling together can provide substantial benefits for real-time VEC systems. Future work will investigate varying network conditions, multi-cell offloading and downloading, and multiple jobs per vehicle. Furthermore, the heuristic will also be improved to compute a schedule in a distributed setting, as there is no need to assign a priority to jobs that do not meet in the pipeline, this will also reduce the runtime. 

\bibliographystyle{IEEEtran}
\bibliography{IEEEabrv,refr}

\end{document}